\documentclass[aps,prb,reprint]{revtex4-1}

\usepackage{graphicx}
\usepackage{amsmath}
\usepackage{amsfonts}
\usepackage{amssymb}
\usepackage{bbm}
\usepackage[]{units}
\usepackage{hyperref}

\usepackage{braket}

\newcommand{\nnb}{\nonumber \\}
\newcommand{\im}{\mathrm{i}}
\newcommand{\bv}{\left( \begin{array}{c}}
\newcommand{\ev}{\end{array} \right)}
\newcommand{\E}{\mathrm{e}}
\newcommand{\ko}{\text{, }}
\newcommand{\po}{\text{. }}
\newcommand{\st}[1]{_{\text{#1}}}
\newcommand{\bo}[1]{\boldsymbol{#1}}
\newcommand{\I}{\mathrm{i}}

\begin{document}
\title{The asymmetric resonant exchange qubit under the influence of electrical noise}
\author{Maximilian Russ and Guido Burkard}
\affiliation{Department of Physics, University of Konstanz, D-78457 Konstanz, Germany}

%%%%%%%%%%%%%%%%%%%%%%%%%%%%%%%%%%%%%%%%%

\begin{abstract}
We investigate the influence of electrical charge noise on a resonant exchange (RX) qubit in a triple quantum dot. This RX qubit is a variation of the exchange-only spin qubit which responds to a narrow-band resonant frequency.  Our noise model includes uncorrelated charge noise in each quantum dot giving rise to two independent (noisy) bias parameters $ \varepsilon $ and $ \Delta $.  We calculate the energy splitting of the two qubit states as a function of these two bias detuning parameters to find ``sweet spots'', where the qubit is least susceptible to noise. Our investigation shows that such sweet spots exist within the low bias regime, in which the bias detuning parameters have the same magnitude as the hopping parameters.  The location of the sweet spots in the $ (\varepsilon,\Delta) $ plane depends on the hopping strength and asymmetry between the quantum dots.  In the regime of weak charge noise, we identify a new favorable operating regime for the RX qubit based on these sweet spots.
\end{abstract}

\maketitle

%%%%%%%%%%%%%%%%%%%%%%%%%%%%%%%%%%%%%%%%%

\section{Introduction}
Universal quantum computing with electron spins in quantum dots \cite{PhysRevA.57.120} has been investigated within a wide range of possible implementations in solid state systems.  Many implementations consider quantum dots in semiconductors, mostly GaAs \cite{Hanson2007,Awschalom2013} or silicon \cite{Zwanenburg2013}. One common feature of these implementations is their dependence on the control of electric and magnetic fields on the nanometer scale which is needed for universal spin control \cite{PhysRevA.57.120}.  This dependence couples the qubits to electric or magnetic noise \cite{Dial2013} introduced by the electric field of the gate voltages, the externally applied magnetic field, and the fluctuating effective magnetic field produced by the nuclear spins.  Although spin coherence times can be fairly long, the susceptibility to electromagnetic noise implies a limitation of the coherence time \cite{Taylor2007}.  Therefore, we are interested in an implementation of a qubit which is protected against noise, in addition to faster qubit control techniques to enable as many coherent operations as possible within the spin lifetime.

Qubits realized with the spin 1/2 of a single electron can be controlled using pulsed magnetic or (via spin-orbit coupling)
electric oscillatory fields, and the exchange interaction can be used to couple such qubits to perform two-qubit gates \cite{PhysRevA.57.120,Petta2005}.
Encoded qubits can be operated with a reduced amount of magnetic or spin-orbit control. 
The encoding into the spin singlet and one of the spin triplet states of two electrons each localized on one of
two nearby quantum dots allows for single qubit rotations generated by the exchange interaction and a static magnetic field gradient \cite{Levy2002,Petta2005,Taylor2005,Foletti2009,Maune2012}. Readout and spin preparation is implemented in singlet-triplet qubits by lowering the potential of one dot and transforming the spin information effectively into a charge signal by means of the Pauli principle.  Two-qubit operations are possible with electrostatic coupling \cite{Taylor2007,Stepanenko2007,Hanson2007,Shulman2012} 
or exchange \cite{Klinovaja2012} between quantum dots belonging to different qubits.

Ultimately, a three electron-spin encoding allows for full qubit control without any magnetic fields and
without relying on the spin-orbit coupling, but with the electrically controllable exchange interaction only  \cite{nature2000,Kawano2005,Fong2011,Zeuch2014}.  
The exchange-only scheme requires the exchange coupling between pairs of spins
to be switched on only for a short period of time.  The fact that the exchange coupling can be switched
off whenever the qubit is idle allows for an advantageos isolation of the spin qubit from the surrounding
charge noise.
The exchange-only qubit can be supplemented with additional control using an oscillatory (typically radio-frequency) electric field when the exchange interaction is constantly turned on.\cite{Laird2010,Medford2013} 
However, this enhanced control comes with additional decoherence
channels due to the effects of charge noise, because the spin singlet and triplet states have slightly
different orbital wavefunctions when the exchange coupling is turned on.
A substantial amount of experimental research has been done on the exchange-only spin qubit implementation since its discovery, e.g. coherent control of the qubit \cite{Gaudreau2012} and readout and qubit preparation \cite{Laird2010}. For readout and spin preparation, techniques used previously for singlet-triplet double quantum dots can be adapted \cite{Laird2010,Medford2013}.
The decoherence produced by the hyperfine interaction with a nuclear bath and by electron-phonon interactions turn out to be of similar magnitude for the subspace and subsystem qubits \cite{Mehl2013}. Here, we consider the influence of charge noise, e.g. from the gate electrodes \cite{Stopa1998} or from impurities in the material \cite{Hayashi2003}. 
It has been predicted that charge noise can indeed be the dominant source of noise.\cite{Hu2006,Culcer2013}

A strength of the exchange qubit with the exchange interactions switched on permanently is the suppression of low-frequency noise, giving rise to a regime in which the system responds to a resonant, narrow frequency band \cite{Medford2013,Taylor2013,Doherty2013}. 
This so called resonant exchange (RX) qubit allows for single-qubit control performed by radio frequency signals in the resonant frequency instead of pulse sequences of the exchange interaction \cite{Medford2013,Taylor2013}. 
Due to this narrow-band response, one can expect a natural protection of the qubit against low frequency electric charge noise. This expectation was confirmed in
the case of one-parameter electric charge noise in the overall energy bias of the triple quantum dot \cite{Taylor2013}.  In this case, a ``sweet spot''  could be identified, 
where the noise is coupled only in second order to the qubit.\cite{Vion2002,Stopa2008,Houck2009}
However, since each of the quantum dots is capacitively coupled to its own electrodes,
and to a different set of charge fluctuators in its immediate vicinity, the electrical potential
on each quantum dot will in reality fluctuate independently, leading to a more general
and more damaging noisy environment than previously assumed.
The effectiveness of sweet spots has recently been investigated for a linear dot geometry in the special case of 
symmetric couplings  $t_l=t_r$, i.e., for $y=0$.\cite{Fei2014} 
In this paper, we address the question whether coherent operation at a sweet spot
of the RX qubit is still possible in these realistic conditions.
To address this question,
we study a noise model with each quantum dot coupled to independent stochastic charge fluctuations.

This paper is organized as follows. In Sec.~\ref{sec:model}, we introduce the model for the qubit under the influence of electrical noise investigated in this paper. 
Subsequently, in Sec.~\ref{sec:dephasing}, the noise model is derived and the dephasing times are calculated. We first show that a sweet spot, if it exists, needs to lie beyond the scope of a perturbative Schrieffer-Wolff (SW) approximation. To go beyond the SW approximation, we also include the low bias regime in our calculations.  To this end, we calculate the spectrum of the exact Hamiltonian and find a sweet spot in a hybridized charge configuration of the triple dot.  In Sec.~\ref{sec:puredephasing}, the pure dephasing times of the qubit at the sweet spot and in the RX regime are calculated in a Ramsey free decay setting. 
We  conclude in Sec.~\ref{sec:outlook} with a summary and an outlook.

%%%%%%%%%%%%%%%%%%%%%%%%%%%%%%%%%%%%%%%%%

\section{Model}
\label{sec:model}

\begin{figure}
	\begin{center}
		\includegraphics[width=1.0\columnwidth]{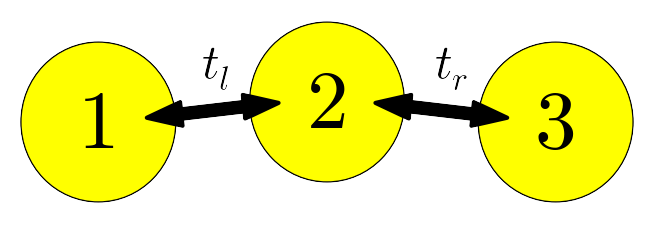}
		\caption{Three quantum dots in an (approximately) linear arrangement, coupled by virtual hopping between the quantum dots 1 and 2 and the quantum dots 2 and 3. Hopping between the first and the third quantum dots is neglected due the nearly linear arrangement. Each quantum dot can be occupied by a maximum of two electrons with opposite spins due the Pauli exclusion principle.}
		\label{fig:arrangement}
	\end{center}
\end{figure}
%\vspace{-10pt}
We consider three quantum dots with a single available orbital state
in a linear arrangement (Fig. \ref{fig:arrangement}),
described by the three-site extended Hubbard Hamiltonian,
\begin{align}
	H\st{Hub}=&\sum_i \left[\frac{U}{2}n_i(n_i-1) + V_in_i\right] + \nnb
	&\sum_{\braket{i,j}}\left[U_cn_in_j + \sum_{\sigma=\uparrow\downarrow}\left(t_{ij}c_{i,\sigma}^\dagger c_{j,\sigma}+\text{h.c.}\right)\right]\ko
	\label{eq:Hub}
\end{align}
with the electron creation and annihilation operators $c_{i,\sigma}^\dagger$ and $c_{i,\sigma}$ on the site (dot) $i$ with spin $\sigma$, the number operator $n_i=\sum_\sigma c_{i,\sigma}^\dagger c_{i,\sigma}$,
the pairwise hopping matrix elements $t_{ij}$ with $i,j\in \{1,2,3\}$ which can be controlled through variation of the gate voltages. Here, we consider symmetric nearest-neighbor hopping ($t_{ij}=t_{ji}$), and set 
$t_{13}=t_{31}=0$, $t_l\equiv\sqrt{2}\, t_{12} = \bar{t} (1-y)$, and $t_r\equiv\sqrt{2}\, t_{23}= \bar{t} (1+y)$.
Here, we have also introduced the mean hopping matrix element $\bar{t}$ and the hopping asymmetry parameter $y$.

The first term in Eq.~(\ref{eq:Hub}) describes the Coulomb energy $U$ required when adding a second electron to one of the dots. The influence of the external gate electrodes is characterized by the second term $V_in_i$, whereas $U_c$ denotes the Coulomb interaction between electrons in nearest-neighbor dots. Filling the triple quantum dot with three electrons allows for 20 charge and spin configurations. There are two states each with a charge configuration $(2,1,0)$, $(2,0,1)$, $(1,2,0)$, $(1,0,2)$, $(0,2,1)$, $(0,1,2)$ and 8 states with $(1,1,1)$, where $(m,n,l)$ denotes a charge state
with $m$ electrons on the left, $n$ in the center, and $l$ in the right dot.
Defining the voltage difference $\varepsilon$ between the outer dots as $\varepsilon\equiv(V_1-V_3)/2$ and the voltage difference $V_m$ between the outer dots and the middle dot as $V_m\equiv(V_3 +V_1 -2V_2)/2$, we note that charge transitions 
between (1,0,2)  and (1,1,1) and between (2,0,1) and (1,1,1)
occur at $\varepsilon=\pm\Delta\equiv\pm(U-2U_c+V_m)$.   

We work in a parameter regime of applied gate voltages $V_i$
where only the charge states (1,1,1), (2,0,1) and (1,0,2) are accessible,
and restrict ourselves to the subspace with total spin $S=1/2$
and spin $z$ projection $S_z=1/2$, spanned by the states\cite{nature2000}
\begin{align}
	&\ket{0}\equiv\ket{s}_{13}\ket{\uparrow}_2=\frac{1}{\sqrt{2}}\left(c_{1,\uparrow}^\dagger c_{2,\uparrow}^\dagger c_{3,\downarrow}^\dagger-c_{1,\downarrow}^\dagger c_{2,\uparrow}^\dagger c_{3,\uparrow}^\dagger\right)\ket{\text{vac}},   \nnb
	&\ket{1}\equiv\!\!\frac{1}{\sqrt{6}}\!\left(2c_{1,\uparrow}^\dagger c_{2,\downarrow}^\dagger c_{3,\uparrow}^\dagger\!-\!c_{1,\uparrow}^\dagger c_{2,\uparrow}^\dagger c_{3,\downarrow}^\dagger\!-\!c_{1,\downarrow}^\dagger c_{2,\uparrow}^\dagger c_{3,\uparrow}^\dagger\right)\!\ket{\text{vac}},  \nnb	
	&\ket{s_{1,1/2}}\equiv\ket{s}_{11}\ket{\uparrow}_3 = c_{1,\uparrow}^\dagger c_{1,\downarrow}^\dagger c_{3,\uparrow}^\dagger\ket{\text{vac}},   \nnb 
	&\ket{s_{3,1/2}}\equiv\ket{\uparrow}_1\ket{s}_{33} = c_{1,\uparrow}^\dagger c_{3,\uparrow}^\dagger c_{3,\downarrow}^\dagger\ket{\text{vac}}\ko
	&\label{eq:logicalspace}
\end{align}
where $\ket{\text{vac}}$ denotes the vacuum state. 
Here, the states $|0\rangle$ and $|1\rangle$ are the logical qubit states
of the exchange-only qubit in the (1,1,1) charge sector, 
while $\ket{s_{1,1/2}}$ and $\ket{s_{1,1/2}}$ denote the accessible states
with the same spin but one doubly occupied quantum dot.
This four-dimensional subspace can be separated from the remaining states by applying a large uniform external magnetic field $B\st{ext}$ such that the states with spin projection $m_s=\pm1/2,\pm3/2$ along the $z$ axis are split by the Zeeman energy. The remaining $ m_s=1/2 $ states have either a total spin $S=1/2$ or $S=3/2$. The states with $S=3/2 $ and charge configuration (1,1,1) are almost completely decoupled from the $S=1/2$ states if the exchange interaction is ongoing and much stronger than the Overhauser field gradients.\cite{Hung2014} Different states than the four defined above can be neglected if one assumes a strong Coulomb repulsion between electrons in neighboring dots (large $ U_C $) and large energy gap between the orbital levels in such a manner that only the lowest orbitals are occupied.
 In the relevant subspace $ \left\lbrace 0,1,s_{1,1/2},s_{3,1/2} \right\rbrace $, the Hamiltonian can be expressed as the $4\times 4$ matrix 
\begin{equation}
	\bar{H}=\left(
\begin{array}{cccc}
 0 & 0 & \left.t_{l}\right/2 & \left.t_{r}\right/2 \\
 0 & 0 & \sqrt{3}\left.t_{l}\right/2 & -\sqrt{3}\left.t_{r}\right/2 \\
 \left.t_{l}\right/2 & \sqrt{3}\left.t_{l}\right/2 & \Delta +\varepsilon  & 0 \\
 \left.t_{r}\right/2 & -\sqrt{3}\left.t_{r}\right/2 & 0 & \Delta -\varepsilon 
\end{array}
\right)\po
\label{eq:hubmatrix}
\end{equation}
\begin{figure}[t]
		\includegraphics[width=1.0\columnwidth]{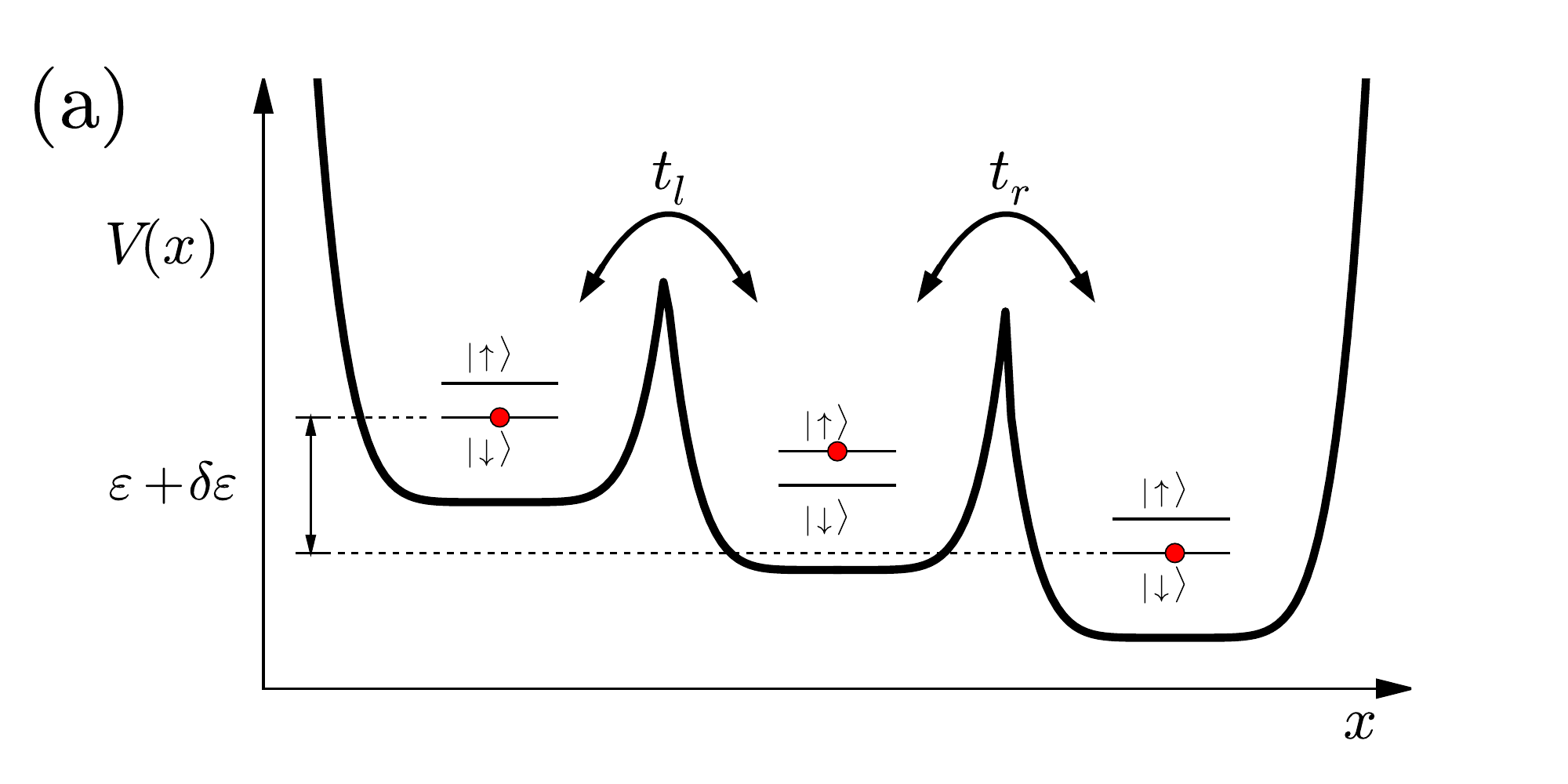}\\
		\includegraphics[width=1.0\columnwidth]{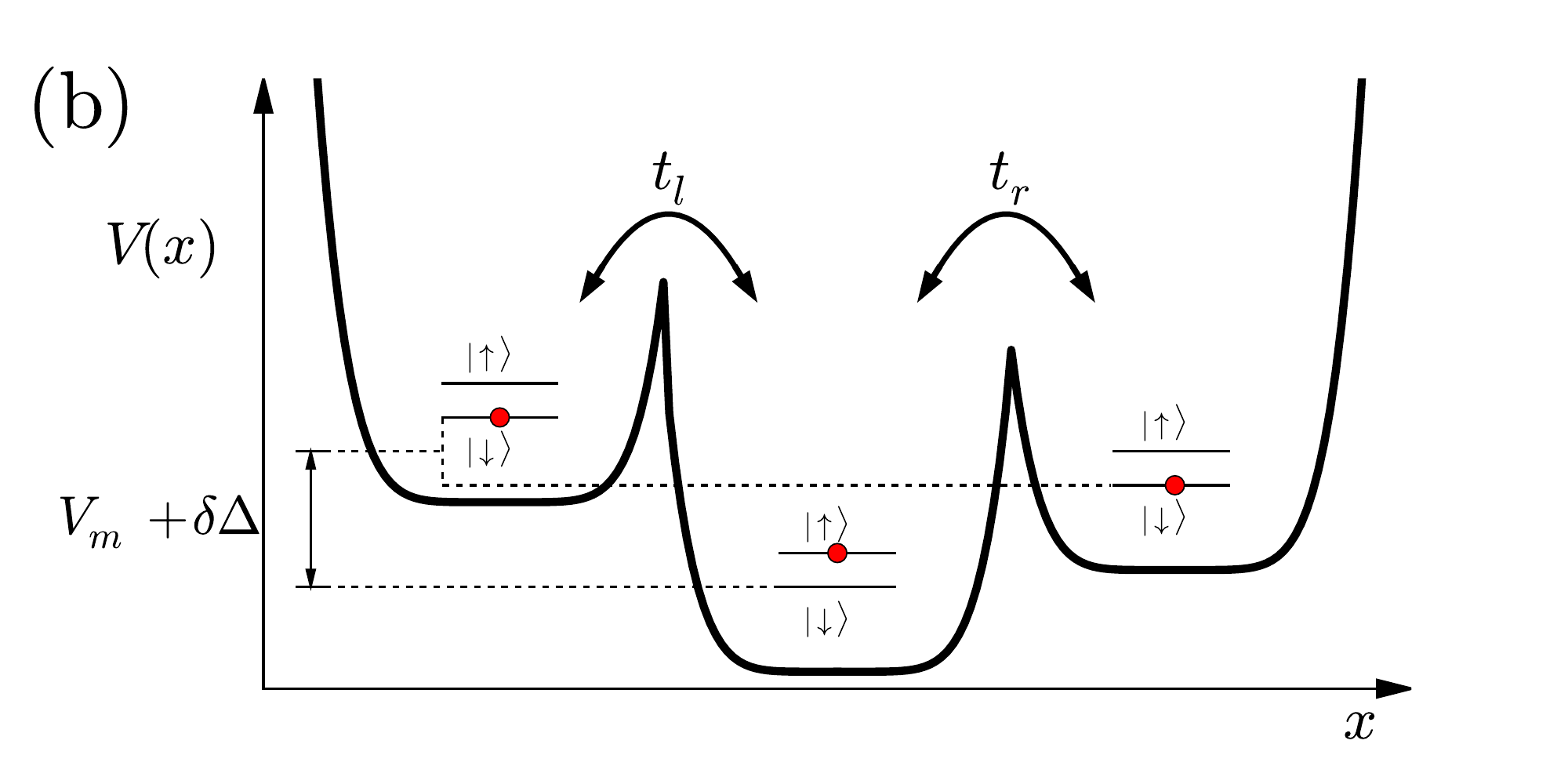}
		\caption{Schematic representation of a triple quantum dot confined by a potential $V(x)$ occupied by three electrons (red dots). There are 20 possible configurations for the electrons if only one orbital per QD is considered. The hopping matrix element of electrons between the quantum dots is denoted by $t_l$ and $t_r$. Here, we describe the response of the system to two noisy energy bias parameters, (a) the energy difference $\varepsilon$  between the outer dots, with noise amplitude $\delta\varepsilon$, and (b) the energy difference  $\Delta$ between the middle dot and the mean energy of the outer dots, with the noise amplitude $\delta\Delta$. \label{fig:noise1}}
\end{figure}

We use a simple model for the description of electrical noise in the gate voltages $V_i$ by adding an independent noise term $\delta V_i$ to each gate voltage, $\tilde{V}_i=V_i+\delta V_i$.  Since only voltage differences affect the relevant spin dynamics, these three noise parameters can be reduced to two parameters. One noise parameter $\delta\varepsilon=(\delta V_1-\delta V_3)/2$ represents noise in the voltage difference between the outer dots $\varepsilon$, as shown in 
Fig.~\ref{fig:noise1}(a). The second parameter is $\delta\Delta=(\delta V_1 +\delta V_3)/2-\delta V_2$ and applies to the center voltage $V_m=(V_3 +V_1 -2V_2)/2$, see Fig.~\ref{fig:noise1}(b).   A third independent noise variable only leads to an inconsequential global phase of the qubit wavefunction.  More precisely, we define the noisy voltage difference $\varepsilon=(\tilde{V}_1-\tilde{V}_3)/2=\varepsilon_0+\delta\varepsilon$ and the noisy effective center voltage $\Delta=U-2U_c - V_{m,0} -\delta\Delta=\Delta_0-\delta\Delta$ which also accounts for fluctuations in the Coulomb energies $U$ and $U_c$. 
Here, the variables with subscript $0$ indicate the parameters in the absence of noise. 

We define the RX qubit as the subspace spanned by the two eigenvectors corresponding 
to the smallest two eigenvalues of $\bar{H}$.  In its eigenbasis, the qubit Hamiltonian can
then be expressed as
 \begin{align}
 	H\st{RX} = \frac{\hbar\omega}{2}\sigma_z ,
 	\label{eq:RXham}
 \end{align}
 with the Pauli $\sigma_z$ matrix and the energy splitting $\hbar \omega$.
The effect of the fluctuations $\delta \varepsilon$ and $\delta\Delta$ in the
parameters $\varepsilon$ and $\Delta$ described by $\bar{H}$ in Eq.~(\ref{eq:hubmatrix})
leads to fluctuating terms in the RX qubit Hamiltonian. In the eigenbasis of the unperturbed 
RX qubit, these terms have the form
\begin{align}
	H\st{RX} = \frac{\hbar}{2} \left( (\omega_0 + \delta\omega_z)\sigma_z + \delta\omega_x\sigma_x 
+ \delta\omega_y\sigma_y \right)
	\label{eq:noiseH}
\end{align}
with the unperturbed eigenfrequency $\omega_0$ and the longitudinal corrections
(up to second order)
\begin{align}
\delta\omega_z  = \omega_{\varepsilon}\delta\varepsilon + \omega_{\Delta}\delta\Delta
+ \frac{\omega_{\varepsilon,\varepsilon}}{2}\delta\varepsilon^2
+ \frac{\omega_{\Delta,\Delta}}{2} \delta\Delta^2
+\omega_{\varepsilon,\Delta} \delta\varepsilon \delta\Delta .
\end{align}
Here, the first derivatives of the qubit frequency
$\omega_{p} = \frac{\partial\omega}{\partial p}\big|_{\varepsilon_0,\Delta_0}$
determine the location of the sweet spot via the condition $\omega_{\varepsilon} = \omega_{\Delta} = 0$ (see below),
while the second derivatives $\omega_{p,q} =  \frac{\partial^2 \omega}{\partial p \partial q}\big|_{\varepsilon_0,\Delta_0}$
with $p,q=\varepsilon,\Delta$ limit the phase coherence of the RX qubit at the sweet spot.
The transverse contributions $\delta\omega_{x,y}$ are needed to calculate the qubit relaxation time, which
is not our concern here.
The longitudinal contribution $\delta\omega_z$ represents the strength of the coupling between the qubit and 
the noise in first order and thus should be eliminated.  Hence, the points $(\varepsilon, \Delta)$ in parameter space
where $\delta\omega_z = 0$  are known as "sweet spots''.
This becomes clear when expanding the eigenenergy difference from Eq.~(\ref{eq:noiseH}),
\begin{eqnarray}
\omega &=& \sqrt{(\omega_0+\delta\omega_z)^2+\delta\omega_x^2+\delta\omega_y^2}\nonumber\\
&\simeq & \omega_0 + \delta\omega_z + \frac{\delta\omega_x^2}{2\omega_0} + \frac{\delta\omega_y^2}{2\omega_0}
+O(\delta\omega^3).
\end{eqnarray}

\begin{figure*}
	\begin{center}
%		\begin{subfigure}[b]{0.45\textwidth}
			\includegraphics[width=0.9\textwidth]{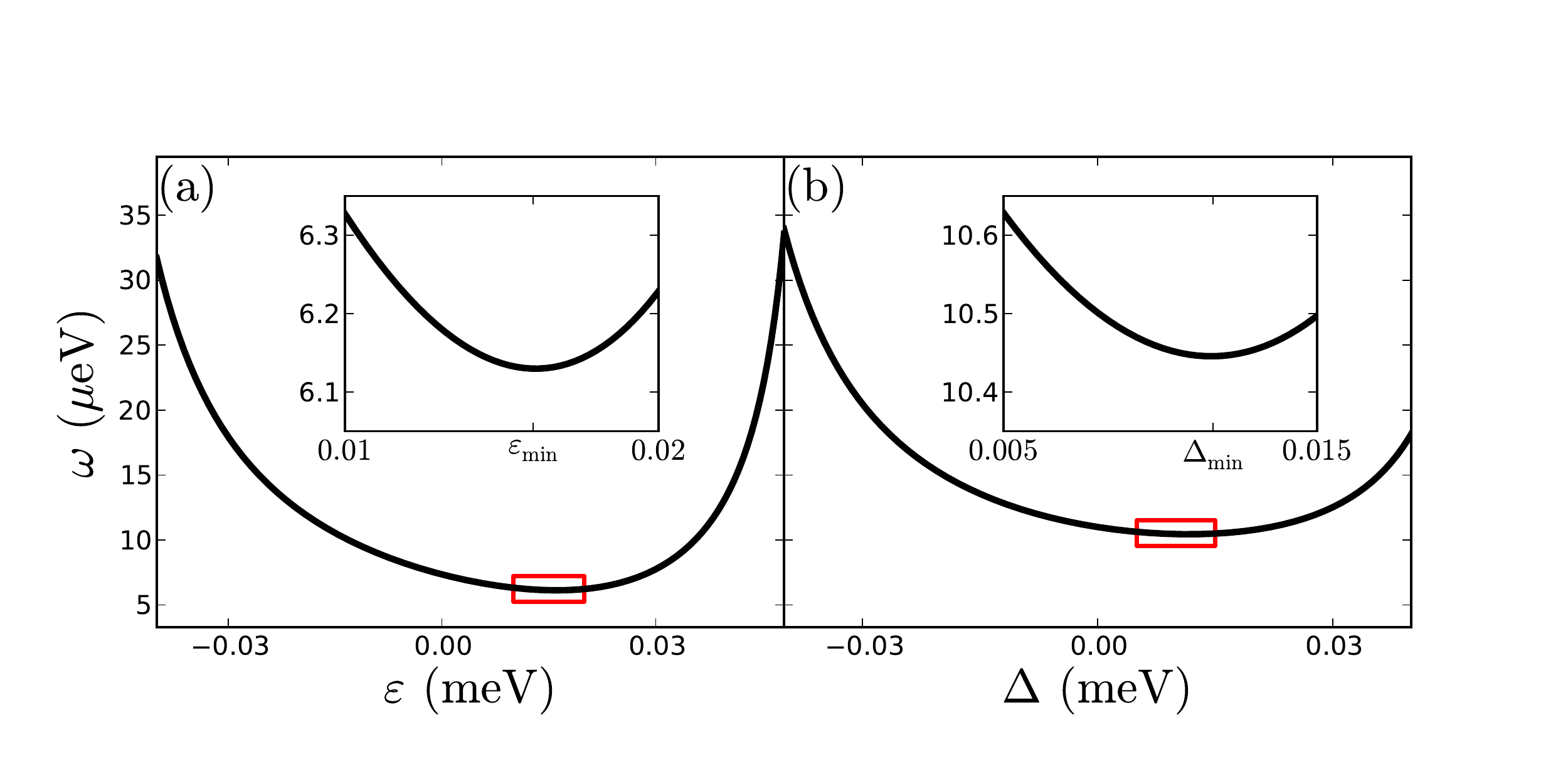}
			\caption{Eigenfrequency $\omega$ as a function of either (a) $\varepsilon_0$ with a minimum at $\varepsilon\st{min}\approx-(8/5)\,y\,\Delta_0$ or (b) $\Delta$ with a minimum at $\Delta\st{min}\approx-(8/7)\,y\,\varepsilon_0$. The parameter $y=-0.1892$ is derived from $t_l=\unit[0.022]{meV}$ and $t_r=\unit[0.015]{meV}$ \cite{Sanchez2014}, while the other parameters are chosen to $\bar{t}=\unit[0.018]{meV}$, in (a) $\Delta=3\bar{t}$ and in (b) $\varepsilon=3\bar{t}$.
The Insets are magnifications of the area around the minima marked with a red rectangle.}
			\label{fig:minimum_e0}	
	\end{center}
\end{figure*}

Away from the degeneracy lines $\varepsilon=\pm \Delta$, the effect of the coupling to the (2,0,1), (1,0,2) states $\ket{s_{1,1/2}}$, $\ket{s_{3,1/2}}$  can be taken into account using an effective Hamiltonian in the low-energy 
(1,1,1) subspace which can
be obtained by applying a Schrieffer-Wolff (SW) transformation $H\st{eff}=\E^{S}\bar{H}\E^{-S}$ such that the resulting matrix is block-diagonal
in lowest order $S \sim t_l, t_r$.  As a result, we find the Heisenberg Hamiltonian,
\begin{align}
	H_{\text{Heis}}=J_l\bo{S}_1\cdot\bo{S}_2 + J_r\bo{S}_2\cdot\bo{S}_3 ,
\end{align}
with the exchange energies $J_l=t_l^2/(\Delta+\varepsilon)$ and $J_r=t_r^2/(\Delta-\varepsilon)$. In the logical subspace spanned by $\ket{0}$ and $\ket{1}$, the Heisenberg Hamiltonian becomes
 \begin{align}
 	H = - \frac{J}{2}\ket{0}\bra{0} -\frac{3J}{2}\ket{1}\bra{1}  - \frac{\sqrt{3}}{2}j \left(\ket{0}\bra{1} + \ket{1}\bra{0}\right)
 	\label{eq:HeisUnter}
 \end{align}
 with the mean exchange parameter $J=(J_l+J_r)/2$ and the exchange difference $j=(J_l-J_r)/2$. 
 Diagonalizing $H$, we can write the RX qubit Hamiltonian in its eigenbasis, Eq.~(\ref{eq:RXham}),
 with
\begin{align}
\hbar \omega=\sqrt{J^2+3j^2}\po
\label{eq:w}
\end{align}

%%%%%%%%%%%%%%%%%%%%%%%%%%%%%%%%%%%%%%%%%

\section{Dephasing of the RX qubit}
\label{sec:dephasing}

\subsection{Non-degenerate regime (SW approximation)}
\label{ssec:nondegenerate}

\begin{figure}[t]
	\begin{center}
		\includegraphics[width=1.0\columnwidth]{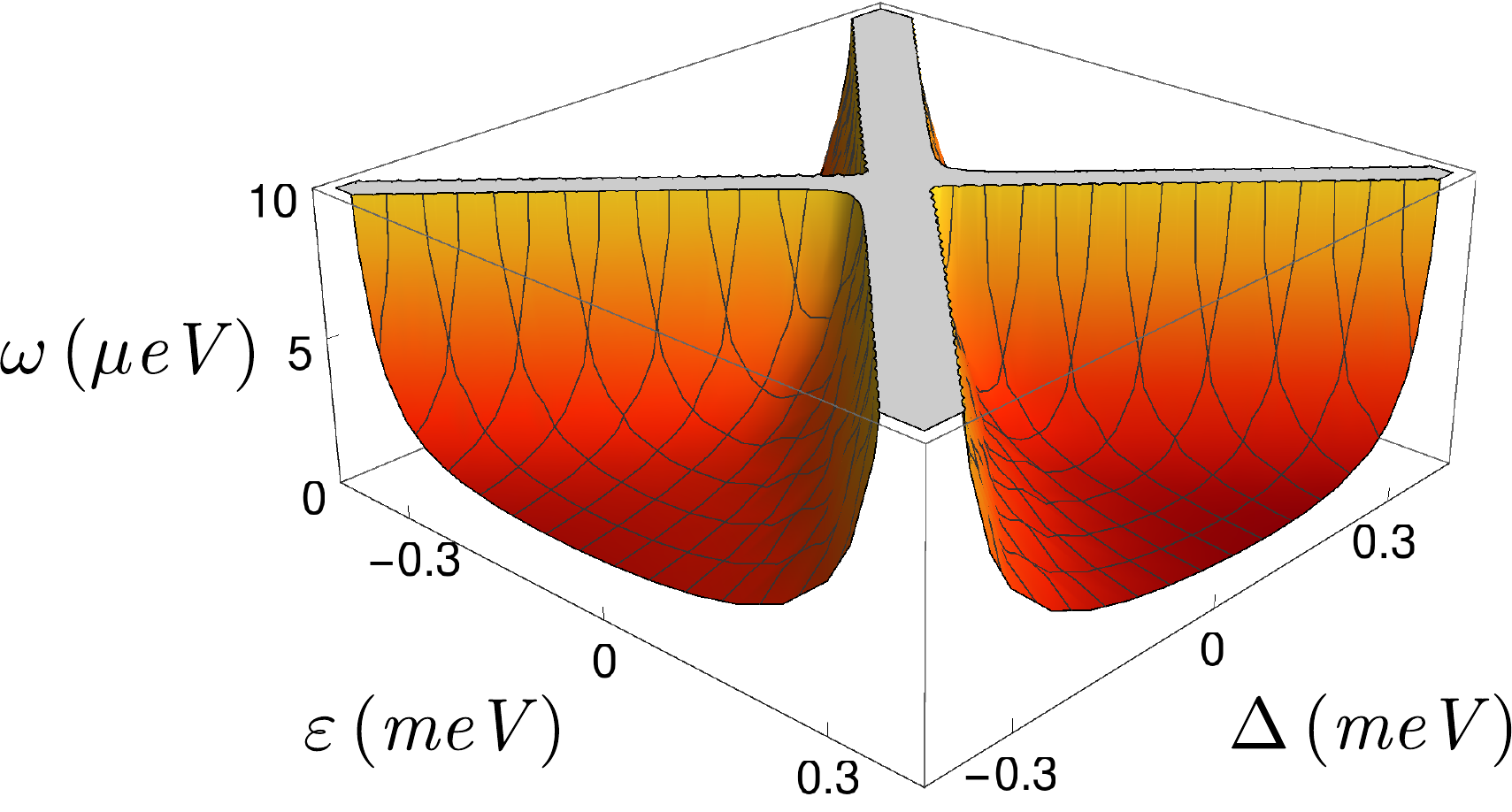}
		\caption{Three dimensional plot of energy gap $\hbar\omega (\Delta,\varepsilon)$. The parameters are chosen to $t_l=\unit[0.022]{meV}$ and $t_r=\unit[0.015]{meV}$ \cite{Sanchez2014}. The clipped parts are diverging and outside the scope of the SW approximation. }
		\label{fig:3dplot}
	\end{center}
\end{figure}

In our analysis, we first investigate the special case of only one noisy detuning parameter, e.g. setting either $\delta\Delta=0$ or $\delta\varepsilon=0$. Our results in this simple case are plotted in 
Fig.~\ref{fig:minimum_e0} and show a minimum of $\omega(\varepsilon,\Delta)$ at (a) $\varepsilon\st{min}\approx-(8/5)\,y\,\Delta_0$ for fixed $\Delta_0$ and (b) $\Delta\st{min}\approx-(8/7)\,y\,\varepsilon_0$ for fixed $\varepsilon_0$, where $ y=(t_r-t_l)/(t_r+t_l) $ denotes the hopping asymmetry.
These minima are sweet spots for one fluctuating parameter; one of them has been studied previously \cite{Taylor2013}.
The qubit energy splitting is in general a function of both $\varepsilon$ and $\Delta$, as shown in Fig.~\ref{fig:3dplot}.

In the non-degenerate regime $|\Delta\pm\varepsilon|\gg t_{l,r}$,
we can study the influence of the electric charge noise on the RX qubit by expanding the low-energy 
Hamiltonian Eq.~\eqref{eq:HeisUnter} to first order in  $\delta\varepsilon$ and $\delta\Delta$.
We transform the result into the eigenbasis of the unperturbed Hamiltonian,
leading us to Eq.~(\ref{eq:noiseH}),
with $\delta \omega_x = \sqrt{3} (J \delta j - j \delta J)/2\omega$, $\delta\omega_y =0$, assuming that
$t_{l,r}$ are real-valued, and $\delta \omega_z = -3 (J \delta J + j \delta j)/2\omega$.
Here, $\delta J = \partial J/\partial\varepsilon|_{\varepsilon_0,\Delta_0} \delta\varepsilon + \partial J/\partial\Delta|_{\varepsilon_0,\Delta_0} \delta\Delta)$, and similarly for $\delta j$.
\begin{figure*}[t]
	\begin{center}
		\includegraphics[width=1\textwidth]{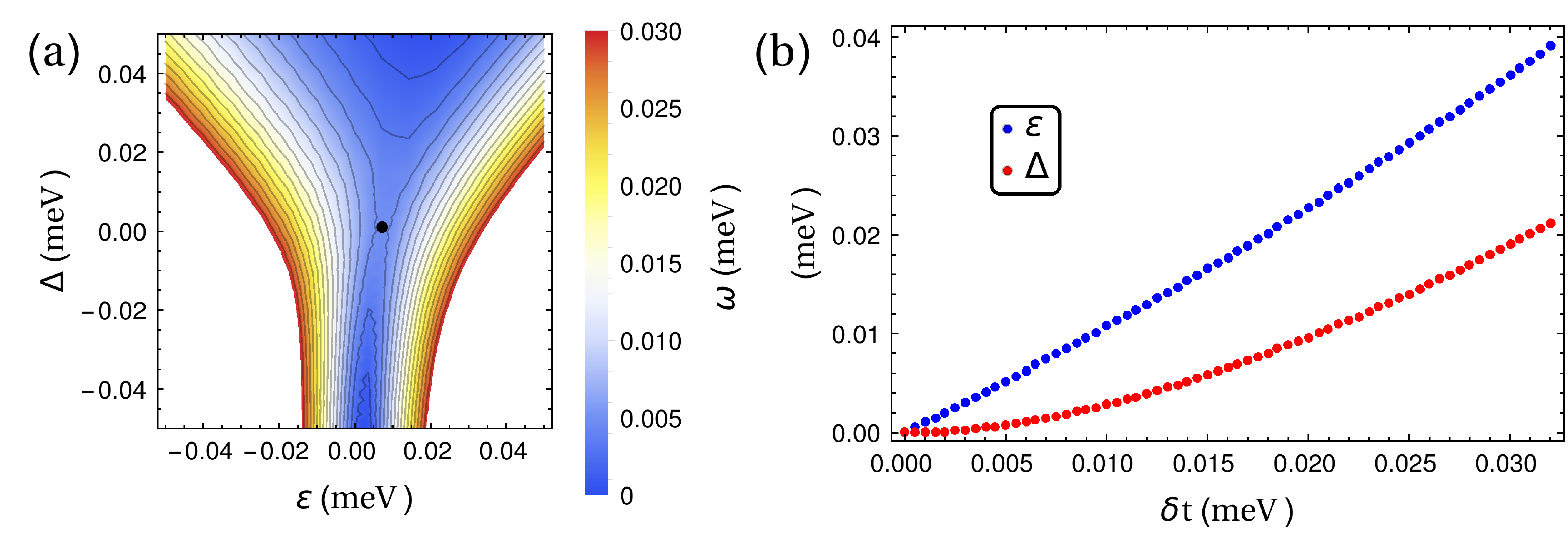}
		\caption{a) Plot of the energy gap between the lowest eigenenergies $\hbar \omega=E_2-E_1$ as a function of the detuning parameters $ \varepsilon,\,\Delta $. We set the hopping parameters to $ t_l=\unit[0.022]{meV} $ and $ t_r=\unit[0.015]{meV} $. The black point marks the position of the sweet spot $ (\unit[0.0074]{meV}, \unit[0.0015 ]{meV})$, which is a saddle point. In b) the position of the sweet spot for the detuning parameters $ \varepsilon $ and $ \Delta $ is plotted as a function of the hopping difference $ \delta t=t_t-t_r $, with $ t_r=\unit[0.015]{meV} $.}	
		\label{fig:energygap}
	\end{center}
\end{figure*}

In the high-bias regime, i.e., away from $\varepsilon=\pm\Delta$, we can expect a sweet spot in the presence of both $\varepsilon$ and $\Delta$ noise if $\delta\omega_z=0$, i.e., if the derivatives of $\omega_z$ with respect to both parameters $\Delta$ and $\varepsilon$ vanish.  We find
\begin{eqnarray}
\frac{\partial\omega}{\partial\varepsilon} 
= \frac{1}{\omega}\left( J\frac{\partial J}{\partial \varepsilon} +3 j\frac{\partial j}{\partial \varepsilon}\right) =\frac{1}{\omega}\frac{1}{\Delta^2-\varepsilon^2}\left(\varepsilon\omega^2-4Jj\Delta\right),\nonumber\\
\\
\frac{\partial\omega}{\partial\Delta} 
= \frac{1}{\omega}\left( J\frac{\partial J}{\partial \Delta} +3 j\frac{\partial j}{\partial \Delta}\right) =\frac{1}{\omega}\frac{1}{\Delta^2-\varepsilon^2}\left(-\Delta\omega^2+4Jj\varepsilon\right),\nonumber\\
\end{eqnarray}
where we have used
\begin{eqnarray}
\frac{\partial J}{\partial\varepsilon} 
=\frac{\partial j}{\partial\Delta} 
=\frac{1}{\omega}\frac{1}{\Delta^2-\varepsilon^2}\left(\varepsilon J-\Delta j\right),\\
\frac{\partial J}{\partial\Delta} 
=\frac{\partial j}{\partial\varepsilon} 
=\frac{1}{\omega}\frac{1}{\Delta^2-\varepsilon^2}\left(-\Delta J+\varepsilon j\right).
\end{eqnarray}
The condition $\frac{\partial\omega}{\partial\varepsilon}=\frac{\partial\omega}{\partial\Delta} =0$ 
cannot be fulfilled for $\varepsilon\neq\Delta$.  Therefore, we cannot find a sweet spot in the
non-degenerate regime.

\subsection{Degenerate regime (exact solution)}
\label{ssec:lowbias}
Since no sweet spot is found within the scope of the SW approximation, we now investigate the degenerate regime
$|\varepsilon \pm\Delta| \lesssim t_{l,r}$,  which is outside the scope of the SW approximation, and  
in particular the low bias regime $\Delta,\varepsilon\rightarrow 0$.
For this purpose, we directly calculate the eigenenergies of the subspace Hubbard Hamiltonian, Eq.~(\ref{eq:hubmatrix}).
We denote the eigenenergies $E_1 \le E_2 \le E_3 \le E_4$ and note that they are functions of the two detuning parameters 
$\varepsilon$ and $\Delta$ as well as the two hopping parameters $t_l$ and $t_r$. 
Analytical expressions for $E_i$ can be obtained, and are shown in Appendix~\ref{ap:exact}.
The qubit states are defined as the two lowest energy levels which match the RX qubits states in the (1,1,1) charge sector,
with energy separation $\hbar\omega = E_2 -E_1$. 
In Fig.~\ref{fig:energygap}(a) we plot $\hbar\omega$ for fixed hopping parameters.
Here, we indeed find a sweet spot (indicated with a black dot) near but not exactly at $\varepsilon=\Delta=0$.  
The position of the sweet spot in $(\varepsilon,\Delta)$ space is shown in 
Fig.~\ref{fig:energygap}(b) for $ t_r=\unit[0.015]{meV} $ as a function of the hopping strength difference $\delta t\equiv t_l-t_r $. 
The formula used to calculate the energy gap and the sweet spot is given in Appendix~\ref{ap:exact}.
The resulting sweet spots always fulfill the condition $\varepsilon \ge \left|\Delta\right|$, hence they are located outside the strict (1,1,1) charge configuration and the qubit states acquire a component of states with a double occupation of the right dot (1,0,2) and the left dot (2,0,1). However, being a sweet spot, the qubit at this working point is only weakly coupled to charge noise. In the special case of symmetric hopping, $ t_l=t_r $, we find a sweet spot at $\varepsilon=\Delta=0$.
One could expect that leakage is problem in this case since the energy gap between the two logical qubit states is comparable to the energy difference to other non-qubit states, whereas in the RX regime the energy gap between the logical qubits is far away from other states.  However, the dynamics show greatly suppressed leakage if only one parameter $\varepsilon$ or $\Delta$ is driven with the resonant frequency $\omega(\varepsilon,\Delta)$.  E.g., Rabi transitions between the qubit states are much faster if 
$\varepsilon$ is periodically driven in contrast to transitions between the energy levels $E_2,\,E_3$, where the detuning parameter $\Delta$ needs to be driven. The reason for the sensitivity to only one parameter is the symmetry of the energy difference with respect to the detuning parameters.  In the fully symmetric case $ t_l=t_r $, Rabi transitions between the energy gaps only occur for one driving parameter, while driving with the other parameter is completely suppressed. For asymmetric hopping the symmetry is weakly broken and Rabi transitions can occur for both parameters but with varying speed.

%%%%%%%%%%%%%%%%%%%%%%%%%%%%%%%%%%%%%%%%%

\section{Pure dephasing}
\label{sec:puredephasing}
\begin{figure*}[t]
	\begin{center}
		\includegraphics[width=1\textwidth]{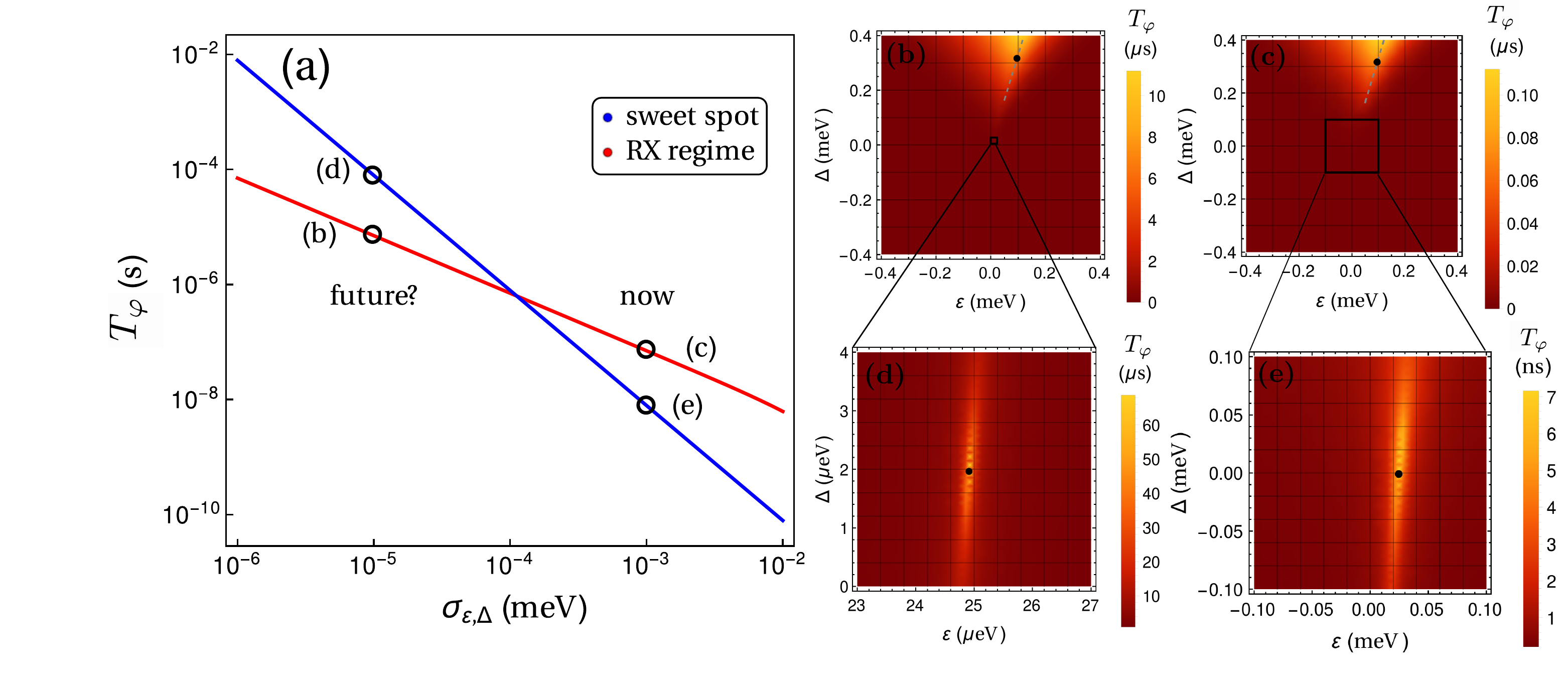}
		\caption{(a) Doubly logarithmic plot of the dephasing time $ T_\varphi $ as a function of the rms (root mean square) value of the noise at the best working points (Sec. \ref{ssec:nondegenerate}) in the RX regime (red) and at the sweet spot (blue). The parameters are chosen as $\Delta= \unit[0.32]{meV}$, $ t_l= \unit[22]{\mu eV}$, $ t_r= \unit[15]{\mu eV}$ and 
$\varepsilon=-(8/5)y\Delta$ for the RX regime plot and $ \varepsilon =\unit[74]{\mu eV}$, $ \Delta =\unit[15]{\mu eV}$, $ t_l= \unit[0.22]{meV}$ and $ t_r=\unit[0.15]{meV}$ for the plot of the sweet spot. (b)-(e) Density plots of dephasing time $ T_\varphi $ as a function of parameter space for different noise amplitudes (black circles in (a)). Note that for the lower plots the hopping parameters are ten times greater than for the upper plots. The black points indicate the position of the best working points plotted in the left (upper) and the sweet spot (lower). The black squares in the upper plots marks the space plotted in the lower ones.}	
		\label{fig:T2noise}
	\end{center}
\end{figure*}

In this section we investigate the effects of charge noise on the RX qubit at the sweet spot in the asymmetric charge configuration (degenerate regime) in comparison to the effects on the RX-qubits\cite{Taylor2013,Medford2013} within the non-degenerate (standard RX regime) with symmetric (1,1,1) charge configuration. Since no sweet spot can be found within the RX regime the eigenenergies couple linearly to both noisy parameters $\delta\varepsilon$ and $\delta\Delta$ giving rise to a dephasing times which scales inversely proportional with the noise amplitude. However within the RX regime, one can find the best working points, where one parameters is minimized, e.g. $\Delta\approx-8/7\,y\,\varepsilon$ or $\varepsilon\approx-(8/5)\,y\,\Delta$, which corresponds to a sweet spot for one parameter, where the dephasing time is inversely proportional in
the other parameter. At the real sweet spot found in this work this scaling is at least inversely quadratic. This characteristic trait can be observed in Fig.~\ref{fig:T2noise} (a), where the estimated dephasing time is plotted as a function of the noise level.  For a current noise level on the order of $ \unit[]{\mu eV} $\cite{Petersson2010,Shi2012}, the RX regime appears to be the better choice, since the resulting dephasing times in the RX regime are two orders of magnitudes longer than at the sweet spots. However, below a noise level on the order of $0.1\,{\mu eV}$, it becomes advantageous to choose the sweet spots due to the better scaling.  In the subsection below, we resent the free decay model used for the calculation of $T_\varphi$.

\subsection{Dephasing model}
\label{ssec:first}
To study dephasing, we start from the noisy RX qubit Hamiltonian Eq.~(\ref{eq:noiseH}) and focus
on the longitudinal noise $\delta \omega_z$.
The time evolution operator $U(t,t_0)$ from an initial time $t_0$ to some later time $t$ can be written as 
\begin{align}
	U(t,t_0)&=\exp\left[-\I\phi(t) \sigma_z\right]
\end{align}
with the accumulated phase 
\begin{align}
\phi(t)&=\int_{t_0}^{t}dt^\prime \delta\omega_z(t^\prime) \nnb
&=\int_{t_0}^{t}dt^\prime \left[
  \omega_{\varepsilon}\delta\varepsilon(t^\prime)
+\omega_{\Delta}\delta\Delta(t^\prime)
+\frac{1}{2}\omega_{\varepsilon,\varepsilon}\delta\varepsilon(t^\prime)^2\right.\nnb
&\left.
+\frac{1}{2}\omega_{\Delta,\Delta}\delta\Delta(t^\prime)^2
+\omega_{\varepsilon,\Delta} \delta\varepsilon(t^\prime)\delta\Delta(t^\prime) \right] \po
\label{eq:phi}
\end{align}
The time ordering operator is not needed, because only longitudinal coupling ($ \sigma_z $) is considered. Considering the effects of an initial coherent superposition of the qubit $\ket{+}=\frac{1}{\sqrt{2}}(\ket{0}+\ket{1})$ we find
\begin{align}
	\ket{\Psi (t)} =\frac{1}{\sqrt{2}}\left[ \ket{0} + e^{\im\phi (t)}\ket{1}\right]\po
	\label{eq:state}
\end{align}
One observable of interest is the mapping on the initial state $P=\ket{+}\bra{+}$ which leads to the free decay ansatz \cite{Taylor2006}
\begin{align}
	\braket{P}=\left| \bra{+}U(t,t_0)\ket{+}\right|^2
	=\frac{1}{2}\left[1+\tilde{f}(t)\right]\po
	\label{eq:mapping}
\end{align}
The function $\tilde{f}(t)$ describes the dephasing in a free decay model and is given for Gaussian distributed noise by
 \begin{align}
 	\tilde{f}(t)&\equiv \left\langle \E^{\I\phi}\right\rangle\approx\exp\left[-\frac{1}{2}\langle\phi(t)^2\rangle\right].
 	\label{eq:decay}
 \end{align}
The detailed formula for the decay and the derivation can be found in Appendix \ref{ap:decay}. For $t\rightarrow\infty$ the superposition is destroyed and the expectation value is $1/2$ for both states, as expected.

\subsection{Approaching real systems}
For further calculations, such as evaluating the integral in Eq.~\eqref{eq:decay}, we require the knowledge of the power spectral density $S(\tilde{\omega})$ of the noise, hence we have to consider electric charge noise in a more detailed manner.  Here, we consider Gaussian distributed noise with a power spectral density $S(\tilde{\omega})=A|\tilde{\omega}|^{-\gamma}$ with variance $ A=\sigma_{\varepsilon,(\Delta)}^2 $ of the noise $ \delta\varepsilon $ ($ \delta\Delta $) and $ \gamma=1 $ which resembles charge noise in double quantum dots.\cite{Dial2013}
The analysis of Eq.~\eqref{eq:decay} leads to Gaussian behavior\cite{Mahklin2004} for the decay rate, $\tilde{f}_l(t)\propto\exp\left[-\left(\frac{t}{T_\varphi}\right)^2+\mathcal{O}(t^3)\right]$, with 
\begin{align}
	T_\varphi=& \hbar \left[ 
  \frac{\omega_{\varepsilon}^2}{2}\,A_\varepsilon\log r
+\frac{\omega_{\Delta}^2}{2}\,A_\Delta\log r \right.\nnb	&+\frac{\omega_{\varepsilon,\varepsilon}^2}{8}A_\varepsilon^2\left(1+2\log r \right)
+\frac{\omega_{\Delta,\Delta}^2}{8}A_\Delta^2\left(1+2\log r\right)\nnb
&+ \left.\frac{2\,\omega_{\varepsilon,\Delta}^2+\omega_{\varepsilon,\varepsilon} \omega_{\Delta,\Delta}}{4}A_\varepsilon\,A_\Delta\right]^{-\frac{1}{2}}
	\label{eq:purepink}
\end{align}
with $r$ as the quotient of upper and lower cutoff, which is needed for ensuring convergence of the integral. The detailed derivation of the formula above can be found in Appendix \ref*{ap:decay}. 
Since at the sweet spot the terms linearly coupled to the noise vanish, one should expect long dephasing times $ T_\varphi $. 
However, our findings shown in Fig.~\ref{fig:T2noise} (e) exhibit shorter dephasing times in comparison with the RX regime 
(Fig.~\ref{fig:T2noise} (c), area with long $T_\varphi$ times) due to a strong contribution of the second order couplings. This contribution, especially $ \omega_{\varepsilon,\varepsilon} $, strongly limits the dephasing time $ T_\varphi $ at a realistic noise level in the order of $ \unit{\mu eV} $\cite{Petersson2010,Shi2012}. Small improvements can be made by considering larger hopping parameters because $\omega_{\varepsilon,\varepsilon}\propto \frac{1}{t_{l,r}}$ or (to a small extent) with a stronger asymmetry, but nonetheless, the dephasing times remain several orders of magnitude shorter than the best points of operation within the RX regime. 
Overall, for the currently available noise level, we find that the sweet spots do not lead to an improvement in coherence and the standard RX regime should be favored instead. 
Those best operation points can be achieved by fine-tuning $\varepsilon$ and $\Delta$ in such a manner that either of the two parameters is minimized (typically $\varepsilon $), while staying within the (1,1,1) charge configuration regime. 
This limits the pure dephasing time to a maximum at $\varepsilon\approx\pm\unit[4]{meV}$ since overly large parameters $\Delta$ and $\varepsilon$ give rise to other charge configurations, effectively reducing the benefit gained by leaving the RX regime. \cite{Taylor2013}
Importantly, the situation changes completely when lower  noise levels become available, because of the quadratic scaling behavior of the dephasing times $ T_\varphi $ at the sweet spot compared to the linear scaling  of $ T_\varphi $ in the RX regime. This different scaling is outlined in Fig.~\ref{fig:T2noise} (a).  For a noise level of $ \unit[10^{-2}]{\mu eV}  $ we find $ T_\varphi $ at the sweet spot one order of magnitude greater than in the RX regime (Fig.~\ref{fig:T2noise} (b) and (d)). The crossover between the two regimes occurs at approximately one order of magnitude less than the currently measured noise leevels. Hence,  by purifying the materials or improving the noise filters in such a manner that the charge noise level is lowered, these sweet spots offer a promising perspective in further  reducing dephasing of charge noise in the future.

%%%%%%%%%%%%%%%%%%%%%%%%%%%%%%%%%%%%%%%%%

\section{Conclusion and Outlook}
\label{sec:outlook}

In this work, we have presented a full description for charge noise in the RX qubit.  We have shown that there are sweet spots for two different coupled noise parameters, which are suppressing charge noise coupled to the RX qubit in first order and give rise to operation points with quadratic noise terms. However, considering both noise parameters simultaneously, no suitable sweet spot is found within the scope of the SW transformation resulting in qubit states linearly coupled to noise. By taking into account the low bias regime, we found a sweet spot outside the scope of the SW approximation in the crossover region to the (2,0,1) and (1,0,2) charge configuration, with a precisely determined location in the $(\varepsilon, \Delta)$ parameter plane depending on the hopping asymmetry.

For the description of the dephasing of the RX qubit we used a Ramsey free decay model to describe the resulting dephasing times $T_\varphi$. We also included quadratic effects which dominate the dephasing at the sweet spots. As a result, we found a Gaussian behavior of the dephasing as a function of time in lowest order. In the next step we compared the usefulness of the sweet spots with the best working points within the RX regime. The resulting analysis shows that the best working points within the RX regime should be favored for currently available noise levels. However, if we consider an improvement by about one order of magnitude in the charge noise level, our sweet spots should be the favored option due to their better scaling behavior of the qubit coherence time.
In this work we have neglected the influence of other noise than charge noise such as spin orbit coupling, hyperfine interaction, \cite{Mehl2013,West2014,Setiawan2014} fluctuations of the homogeneous magnetic field and so on.  In future studies, these effects can be included
in a full quantum master-equation approach to further improve the results.

%%%%%%%%%%%%%%%%%%%%%%%%%%%%%%%%%%%%%%%%%

\section*{Acknowledgments}
We thank Niklas Rohling for helpful discussions. This work was supported by Deutsche Forschungsgemeinschaft (DFG) through SFB767 and the European Union throught the Marie Curie ITN S$^3$NANO.

%%%%%%%%%%%%%%%%%%%%%%%%%%%%%%%%%%%%%%%%%

\appendix

%%%%%%%%%%%%%%%%%%%%%%%%%%%%%%%%%%%%%%%%%

% \section{First derivative test}
% \label{ap:first}
% The first derivative test for the eigenfrequency from Eq. \eqref{eq:w} leads to the following two equations:
% \begin{widetext}
% \begin{align}
% 	-(\Delta_0-\varepsilon_0)^2 (-1+y)^2 \left\lbrace\varepsilon_0 \left[-3+(2-3 y) y\right]+\Delta_0 \left[1+(-6+y) y\right]\right\rbrace =& 0 \\
% 	-(\Delta_0 + \varepsilon_0)^2 (1 + y)^2 \left\lbrace\Delta_0 \left[1 + y (6 + y)\right] + \varepsilon_0 \left[3 + y (2 + 3 y)\right] \right\rbrace =& 0
% \end{align}
% \end{widetext}
% With the requirements $|y|<1$ (the case $y=\pm 1$ signalizes no coupling between dot 1 and 2 or 2 and 3) and $\Delta_0\neq \pm \varepsilon_0$ (leads to a diverging eigenenergy $\omega$ ) we gain after straight forward calculus
% \begin{align}
% 	\Delta_0\frac{1+y^2}{2y}&=\varepsilon_0\ko\\
% 	\Delta_0\,(1+y^2)\left(\frac{1}{2y}+0,5y+1\right) &=0\po
% \end{align}
% The equation system has only the trivial solution $\varepsilon_0=\Delta_0=0$, which is not within the scope of the SW approximation.

%%%%%%%%%%%%%%%%%%%%%%%%%%%%%%%%%%%%%%%%%

\section{Exact solution of the subspace Hubbard Hamiltonian}
\label{ap:exact}
To describe the parameter space for small $\varepsilon,\Delta$, which is outside the scope of the SW approximation, we calculate the eigenenergies of the Hubbard Hamiltonian in the subspace spanned by $\left\lbrace\ket{0},\ket{1},\ket{s_{1,1/2}},\ket{s_{3,1/2}}\right\rbrace$, Eq.~\eqref{eq:hubmatrix}, directly with the general solution for a polynomial of order four. This is giving rise to the four eigenenergies (numbered from lowest to highest)
\begin{align}
	E_{1}&= \frac{\Delta}{2}-\frac{\eta}{2}-\frac{\kappa_-}{2},\nnb
	E_{2}&= \frac{\Delta}{2}-\frac{\eta}{2}+\frac{\kappa_-}{2},\nnb
	E_{3}&= \frac{\Delta}{2}+\frac{\eta}{2}-\frac{\kappa_+}{2},\nnb
	E_{4}&= \frac{\Delta}{2}+\frac{\eta}{2}+\frac{\kappa_+}{2},
	\label{eq:exacteig}
\end{align}
with the abbreviations
%\begin{widetext}
\begin{align}
	\kappa_\pm\equiv&\sqrt{\Delta^2+\varepsilon^2+t_l^2+t_r^2 -\frac{\alpha}{3}-\frac{\gamma}{3\sqrt[3]{2}} \pm\frac{2\Delta^3-\Delta\alpha-\beta}{\eta}},\nnb
	\eta \equiv& \sqrt{\varepsilon^2+t_l^2+t_r^2+\frac{\alpha}{3}+\frac{\gamma}{3\sqrt[3]{2}}+\zeta},\nnb
	\zeta \equiv& \frac{\sqrt[3]{2}\left(\alpha^2+6\Delta\beta+9\,t_l^2\,t_r^2\right)}{3\gamma},\nnb
	\gamma \equiv& \Big(  \left\lbrace\left[27 \left(3 \Delta^2 \,t_l^2 \,t_r^2+\beta^2\right)+18 \alpha \left(\Delta \beta-3\, t_l^2\,t_r^2\right)+2 \alpha^3\right]^2 \right.\nnb
	&\left. -4 \left(6 \Delta \beta+\alpha^2+9\, t_l^2\, t_r^2\right)^3  \right\rbrace^{1/2} \nnb
	&+81 \Delta^2\, t_l^2 \,t_r^2+18 \Delta \alpha \beta+2 \alpha^3-54\, \alpha \,t_l^2\, t_r^2+27 \beta^2\Big)^{1/3},\nnb
	\beta \equiv& \left(\Delta - \varepsilon\right) t_l^2 + \left(\Delta + \varepsilon\right) t_r^2,\nnb
	\alpha \equiv& \Delta^2 -\varepsilon^2 - t_l^2 -t_r^2\po
	\label{eq:abbr}
\end{align}
%\end{widetext}
In our investigation we only consider the energy gap between the lowest two levels $\omega\equiv E_1-E_2=\kappa_- $ and its partial derivatives $ \frac{\partial\kappa_-}{\partial\varepsilon} $ and $ \frac{\partial\kappa_-}{\partial\Delta} $. The condition for the sweet spot $ \frac{\partial\kappa_-}{\partial\varepsilon} = \frac{\partial\kappa_-}{\partial\Delta} =0 $ is giving rise to a system of nonlinear equations. Instead we minimized the equivalent system
\begin{align}
\left(\frac{\partial\kappa_-}{\partial\varepsilon}\right)^2 + \left(\frac{\partial\kappa_-}{\partial\Delta}\right)^2 = 0\ko
\label{eq:conditionsweet}
\end{align}
using that all parameters are real valued.

%%%%%%%%%%%%%%%%%%%%%%%%%%%%%%%%%%%%%%%%%

\section{Derivation of the free decay rates}
\label{ap:decay}
For an estimation of the dephasing rate, the observable of interest is the projection on the initial state Eq.~\eqref{eq:mapping}. Therefore, it is sufficient to calculate 
\begin{align}
	\tilde{f}(t)\equiv \left\langle e^{\im\phi}\right\rangle\ko
	\label{eq:dec}
\end{align}
which can be expanded by using the cumulants 
\begin{align}
	\log\left\langle e^{\im\phi}\right\rangle = \im \left\langle \phi (t)\right\rangle_1^c -\frac{1}{2} \left\langle \phi(t)\right\rangle_2^c + \cdots.
\end{align}
Assuming Gaussian noise with zero mean $\left\langle \delta\varepsilon(t) \right\rangle=\left\langle \delta\Delta(t) \right\rangle=0$, all cumulants higher than two and all odd moments vanish, resulting in
\begin{widetext}
\begin{align}
 \log\left\langle e^{\im\phi}\right\rangle =-\frac{1}{2}\left\langle\phi(t)^2\right\rangle %\nnb
 =&- \frac{1}{2}\left( \left\langle\left(\int_{0}^{t} \omega_{\varepsilon}\delta\varepsilon(t^\prime)dt^\prime \right)^2\right\rangle %\right.\nnb
%&\left. 
+\left\langle\left(\int_{0}^{t} \omega_{\Delta}\delta\Delta(t^\prime)dt^\prime \right)^2\right\rangle
+\frac{1}{4}\left\langle\left(\int_{0}^{t} \omega_{\varepsilon,\varepsilon}\delta\varepsilon(t^\prime)^2dt^\prime \right)^2\right\rangle
\right.\nnb &\left. 
+ \frac{1}{4}\left\langle\left(\int_{0}^{t} \omega_{\Delta,\Delta}\delta\Delta(t^\prime)^2 dt^\prime \right)^2\right\rangle
%\right.\nnb &\left.
+\frac{1}{2} \left\langle\left(\int_{0}^{t} \omega_{\varepsilon,\Delta}^2\delta\varepsilon(t^\prime)^2\delta\Delta(t^\prime)^2 dt^\prime \right)\right\rangle
\right.\nnb &\left. 
+\frac{1}{2}\left\langle\left(\int_{0}^{t} \omega_{\varepsilon,\varepsilon}\delta\varepsilon(t^\prime)^2dt^\prime \right)\left(\int_{0}^{t} \omega_{\Delta,\Delta}\delta\Delta(t^\prime)^2 dt^\prime \right)\right\rangle\right)
\end{align}

Fourier transforming and some calculus leads to \cite{Ithier2005,Taylor2006}
%\begin{widetext}
\begin{align}
	\log\left\langle e^{\im\phi}\right\rangle &=-\frac{t^2\,\omega_{\varepsilon}^2}{2} \int_{-\infty}^{\infty}d\tilde{\omega}\, S_\varepsilon(\tilde{\omega}) \,\text{sinc}^2\left(\frac{\tilde{\omega}t}{2}\right)-\frac{t^2\,\omega_{\Delta}^2}{2} \int_{-\infty}^{\infty}d\tilde{\omega}\, S_\Delta(\tilde{\omega}) \,\text{sinc}^2\left(\frac{\tilde{\omega}t}{2}\right) \nnb
	&-\frac{\omega_{\varepsilon,\varepsilon}^2t^2}{8}\left(\left\langle\delta\varepsilon^2\right\rangle^2+2\iint_{-\infty}^\infty d\omega\,d\omega^\prime\text{sinc}^2\left[\left(\frac{\omega + \omega^\prime}{2}\right) \right] S_\varepsilon(\omega)S_\varepsilon(\omega^\prime)\right)-\frac{\omega_{\varepsilon,\varepsilon}\omega_{\Delta,\Delta}t^2}{4}\left\langle\delta\varepsilon^2\right\rangle\left\langle\delta\Delta^2\right\rangle\nnb &-\frac{\omega_{\Delta,\Delta}^2t^2}{8}\left(\left\langle\delta\Delta^2\right\rangle^2+2\iint_{-\infty}^\infty d\omega\,d\omega^\prime\text{sinc}^2\left[\left(\frac{\omega + \omega^\prime}{2}\right) \right] S_\Delta(\omega)S_\Delta(\omega^\prime)\right)-\frac{\omega_{\varepsilon,\Delta}^2t^2}{2}\left\langle\delta\varepsilon^2\right\rangle\left\langle\delta\Delta^2\right\rangle ,
 	\label{eq:decayrate}
\end{align}
\end{widetext}
where $S_q(\omega)=\int_{-\infty}^\infty \delta q(\tau) \E^{-\im\omega \tau}d\tau$ ($ q\in \lbrace\varepsilon,\Delta\rbrace $) denotes the spectral energy density of the noise coupling to bias parameter $ q $.

\subsection{First Order}
 For our investigation we consider noise with a spectral density  $S(\tilde{\omega})=\frac{A}{|\tilde{\omega}|}$, which is anti-proportional to the frequency. Set into equation \eqref{eq:decayrate} yields the free decay
\begin{align}
	\tilde{f}(t)&=\exp\left[-\frac{t^2\,\omega_{q}^2\,A}{2} \int_{-\infty}^{\infty} d\tilde{\omega}\,\, \frac{A_q}{|\tilde{\omega}|} \text{sinc}^2\left(\frac{\tilde{\omega}t}{2}\right)\right],
\end{align}
which diverges at the lower limit. To ensure convergence of the integral the spectral density is modified to cutoff the lowest ($\tilde{\omega}\leq\omega_R$) and highest frequencies ($\tilde{\omega}\geq\omega_U$) with $\omega_U>\omega_R>0$, hence $S(\tilde{\omega})=\frac{A}{|\tilde{\omega}|} \Theta(\tilde{\omega}-\omega_R) \Theta(\omega_U-\tilde{\omega}) $. This modification leads to

\begin{align}
	\tilde{f}_1(t)\equiv\exp\left[-2\,t^2\,\,\omega_{q}^2\,A_q \int_{\omega_R}^{\omega_U} d\tilde{\omega}\,\, \frac{A}{\tilde{\omega}^3} \sin^2\left(\frac{\tilde{\omega}t}{2}\right)\right],
\end{align}
which can be integrated to
\begin{widetext}
\begin{align}
	\tilde{f}_1(t)=\exp\left[- 2\,\omega_{q}^2\,A_q \left(-\frac{-1+\cos(t \omega_R)+t^2 \omega_R^2 \text{ci}(t \omega_R)-t \omega_R \sin(t \omega_R)}{4 \omega_R^2} \right. \right. \nnb
	\left.\left. + \frac{-1+\cos(t \omega_U)+t^2 \omega_U^2 \text{ci}(t \omega_U)-t \omega_U \sin(t \omega_U)}{4 \omega_U^2}\right)\right],
	\label{eq:sol}
\end{align}
with cosine integral $\text{ci}(x)= -\int_x^\infty\frac{\cos t}{t}\,dt$.
In the limit $\omega_U\rightarrow\infty$, all high frequencies are valued, the second term in the exponent vanishes and the solution is

\begin{align}
	\tilde{f}_{P,\infty}(t)=\exp\left[- 2\,\omega_{q}^2\,A_q \left(-\frac{-1+\cos(t \omega_R)+t^2 \omega_R^2 \text{ci}(t \omega_R)-t \omega_R \sin(t \omega_R)}{4 \omega_R^2}\right)\right].
\end{align}
Considering small $t \omega_R\ll 1$, the first order of the evolution of the exponent can be written as
\begin{align}
	\tilde{f}_{1,\infty}(t)\approx\exp\left[- \frac{t^2\,\omega_{q}^2\,A_q}{2}\left(\log\left(\frac{1}{\omega_R\,t}\right) + \underbrace{(1 - \gamma)}_{\approx 0\text{.}42}\right)\right]\text{.}
\end{align}
Since only the lower frequency regime is dominated by pink noise and pink noise can be neglected for higher frequencies, a high frequency cutoff can be motivated and the two terms in equation \eqref{eq:sol} can be simplified to
\begin{align}
	\tilde{f}_{1}(t)=\exp\Big[-2\,\omega_{q}^2\,A_q \bigg(\underbrace{\frac{1}{4}\left( \frac{1}{\omega_R^2}-\frac{1}{\omega_U^2} \right) }_{\equiv \mathrm{const}>0}+ \frac{t^2}{4}\left( \text{ci} ( \omega_U\,t)-\text{ci} ( \omega_R\,t)\right)
	+\frac{t^2}{4}\left(\text{sinc}(\omega_R\,t)-\text{sinc}(\omega_U\,t)\right)\bigg)\Big]\text{.}
\end{align}
Considering both $t \omega_R\ll 1$ and $t \omega_U\gg 1$ to be small, the exponent can be expanded and rewritten as \cite{Ithier2005}
\begin{align}
	\tilde{f}_{1}(t)\approx \exp\Big[-2\,t^2\,\,\omega_{q}^2\,A_q \bigg(k + \frac{t^2}{4}\log\left(\frac{\omega_U}{\omega_R}\right)\bigg)\Big]\propto 
	 \exp\left[-\omega_{q}^2\,A_q \frac{t^2}{4}\log\left(\frac{\omega_U}{\omega_R}\right)\right]\text{.}
	 \label{eq:pinknoise}
\end{align}

Introducing $r = \frac{\omega_U}{\omega_R}$ the expression above takes the form of the first two terms in Eq. \eqref{eq:purepink}.
\subsection{Second Order}
For the second order terms we have to calculate
\begin{align}
\iint_{-\infty}^\infty d\omega\,d\omega^\prime\text{sinc}^2\left[\left(\frac{\omega + \omega^\prime}{2}\right) \right] S_{\varepsilon,\Delta}(\omega)S_{\varepsilon,\Delta}(\omega^\prime),
\end{align}
from Eq. \eqref{eq:decayrate} which can be done in a similar manner than the linear term above by introducing  the same power spectral density $S(\tilde{\omega})=\frac{A}{|\tilde{\omega}|} \Theta(\tilde{\omega}-\omega_R) \Theta(\omega_U-\tilde{\omega}) $ with a low and high frequency cutoff. The integration gives 
\begin{align}
\tilde{f}_{2}(t) =& \frac{\omega_{q,q}^2t A_q^2}{4\omega_U \omega_R}\left\lbrace t \omega_U \omega_R \text{ci}\left(\frac{t \omega_U}{2}\right)^2+t \omega_U \omega_R \text{ci}\left(\frac{t \omega_R}{2}\right)^2-2 \text{ci}\left(\frac{t \omega_U}{2}\right) \left[t \omega_U \omega_R \text{ci}\left(\frac{t \omega_R}{2}\right)+2 \omega_R \sin \left(\frac{t \omega_U}{2}\right) \right.\right.\nnb
&\left.\left.-2 \omega_U \sin \left(\frac{t \omega_R}{2}\right)\right]+4 \text{ci}\left(\frac{t \omega_R}{2}\right) \left[\omega_R \sin \left(\frac{t \omega_U}{2}\right)-\omega_U \sin \left(\frac{t \omega_R}{2}\right)\right]-t \omega_U \omega_R \text{si}\left(\frac{t \omega_U}{2}\right)^2\right.\nnb
&\left.-t \omega_U \omega_R \text{si}\left(\frac{t \omega_R}{2}\right)^2+2 t \omega_U \omega_R \text{si}\left(\frac{t \omega_U}{2}\right) \text{si}\left(\frac{t \omega_R}{2}\right)+4 \omega_R \text{si}(t \omega_U)+4 \omega_U \text{si}(t \omega_R)\right.\nnb
&\left.- 4 (\omega_U+\omega_R) \text{si}\left(\frac{1}{2} t (\omega_U+\omega_R)\right)-4 \omega_R \text{si}\left(\frac{t \omega_U}{2}\right) \cos \left(\frac{t \omega_U}{2}\right)+4 \omega_U \text{si}\left(\frac{t \omega_U}{2}\right) \cos \left(\frac{t \omega_R}{2}\right)\right.\nnb
&\left.+4 \omega_R \text{si}\left(\frac{t \omega_R}{2}\right) \cos \left(\frac{t \omega_U}{2}\right)-4 \omega_U \text{si}\left(\frac{t \omega_R}{2}\right) \cos \left(\frac{t \omega_R}{2}\right)\right\rbrace\ko
\end{align}
 but can be approximated by a Taylor series expansion into the familiar expression
\begin{align}
\tilde{f}_{2}(t)\approx\exp\left\lbrace-\frac{\omega_{q,q}^2t^2 A_q^2}{4} \left[\log^2\left(\frac{\omega_U}{\omega_R}\right)+\mathcal{O}(t^2)\right]\right\rbrace.
\end{align}
With $r = \frac{\omega_U}{\omega_R}$ the expression above takes the form of the missing two terms in Eq. \eqref{eq:purepink}.
\end{widetext}

\bibliography{lit}

\end{document}